\documentclass[12pt]{article}

\usepackage{amsmath,amssymb,amsfonts,graphics,graphicx,amscd,amsfonts,epsfig,color}
\usepackage{subfig}
\usepackage{here}
\newcommand {\beq} {\begin{equation}}
\newcommand {\eeq} {\end{equation}}
\newcommand {\beqa}{\begin{eqnarray}}
\newcommand {\eeqa}{\end{eqnarray}}

\newcommand {\tr}{{\rm tr\,}}

\makeatletter
    
    \@addtoreset{equation}{section}
  \makeatother

\allowdisplaybreaks[4]

\date{}
\begin{document}

\begin{flushright} 
KEK-TH-1952
\end{flushright} 

\vspace{1.0cm}

\begin{center}
{\bf \large Universality 
and the dynamical space-time dimensionality
in the Lorentzian type IIB matrix model}
\end{center}

\vspace{1.5cm}

\begin{center}

         Yuta I{\sc to}$^{a}$\footnote
          {
 E-mail address : yito@post.kek.jp},  
         Jun N{\sc ishimura}$^{ab}$\footnote
          {
 E-mail address : jnishi@post.kek.jp} 
and
         Asato T{\sc suchiya}$^{c}$\footnote
          {
 E-mail address : tsuchiya.asato@shizuoka.ac.jp} 

\vspace{0.5cm}

$^a${\it KEK Theory Center, 
High Energy Accelerator Research Organization,\\
1-1 Oho, Tsukuba, Ibaraki 305-0801, Japan} 

$^b${\it Graduate University for Advanced Studies (SOKENDAI),\\
1-1 Oho, Tsukuba, Ibaraki 305-0801, Japan} 

$^c${\it Department of Physics, Shizuoka University,\\
836 Ohya, Suruga-ku, Shizuoka 422-8529, Japan}

\end{center}

\vspace{1.5cm}

\begin{center}
  {\bf abstract}
\end{center}

\noindent The type IIB matrix model
is one of the most promising candidates
for a nonperturbative formulation of superstring theory.
In particular, its Lorentzian version was shown to exhibit an interesting
real-time dynamics such as the spontaneous breaking of the 9-dimensional
rotational symmetry to the 3-dimensional one.
This result, however, was obtained after regularizing the original
matrix integration by introducing ``infrared'' cutoffs on 
the quadratic moments of the Hermitian matrices.
In this paper, we generalize the form of the cutoffs in such a way that
it involves an arbitrary power ($2p$) of the matrices.
By performing Monte Carlo simulation of a simplified model,
we find that the results become independent of $p$
and hence universal for 
$p \gtrsim 1.3$.
For $p$ as large as 2.0, however,
we find that large-$N$ scaling behaviors
do not show up, and we cannot take a sensible large-$N$ limit.
Thus we find that there is a certain range of $p$
in which a universal large-$N$ limit can be taken.
%
Within this range of $p$,
the dynamical space-time dimensionality 
turns out to be
$(3+1)$, while 
for $p=2.0$, where we cannot take a sensible large-$N$ limit,
we observe a (5+1)d structure.


\newpage

\section{Introduction}

Since its proposal in 1996 \cite{Ishibashi:1996xs}, 
the type IIB matrix model has been studied from various angles
as a possible nonperturbative formulation of 
superstring theory.
Formally it resembles the proposals 
in refs.~\cite{Banks:1996vh,Dijkgraaf:1997vv}
in the sense that all the models can be obtained
by dimensionally reducing 10D $\mathcal{N}=1$ super Yang-Mills action
to lower dimensions.
From this point of view, the type IIB matrix model may be viewed as
an extreme case since the dimensional reduction is conducted down 
to $d=0$.
This makes the model distinct in that
not only space but also time is treated as an emergent concept that
appears from the matrix degrees of freedom,
and consequently the model enjoys manifest Lorentz invariance.
There is also a strong evidence 
that the model can reproduce the perturbation 
theory of type IIB superstring theory to all orders 
in the string coupling constant \cite{Fukuma:1997en}.

In the literature, the type IIB matrix model was studied mostly
after making a ``Wick rotation'' $A_0 = i A_{10}$, where $A_0$ represents
the matrix corresponding to the time.
The Euclidean version obtained in this way has 
a positive semi-definite action for the bosonic part,
and the partition function is proved to be 
finite in spite of the existence of 
flat directions \cite{Krauth:1998xh,Austing:2001pk}.
The SO(10) symmetry of the model is expected to be
spontaneously broken down to SO(4)
in order to realize 
the dynamical generation of four-dimensional space-time \cite{Aoki:1998vn}.
The latest result obtained by the Gaussian expansion method suggests,
however, that it is broken down to SO(3), and the extent of space in the
extended directions is only five times larger than 
the shrunken directions \cite{Nishimura:2011xy}.

The Lorentzian version of the type IIB matrix model,
on the other hand, 
is not well-defined as it is since 
the bosonic part of the action is not positive semi-definite.
In ref.~\cite{Kim:2011cr},
the model was studied by Monte Carlo simulation
after regularizing the matrix integral by introducing
``infrared'' cutoffs on $\tr (A_0)^2$ and $\sum_{i=1}^{9} \tr (A_i)^2$
for the temporal and spatial directions, respectively.
The matrix configurations obtained by the simulation were found to have 
an approximate band-diagonal structure in the basis which diagonalizes
the temporal matrix as $A_0 = {\rm diag} (\alpha_1 , \cdots , \alpha_N)$
with the order $\alpha _ 1 < \cdots < \alpha_N$. 
Therefore it makes sense to
identify the state at time $\alpha_a$
by cutting out a block matrix from $A_i$ ($i=1, \cdots , 9$)
around the diagonal element at the position of $\alpha_a$ in $A_0$.
The real-time evolution extracted in this way
showed the following exciting behavior:
the extent of space in three out of nine directions starts to grow
at some point in time
indicating that the SO(9) rotational symmetry of the spatial matrices
is broken down to SO(3) at that point.
The expanding behavior is speculated to be 
exponential at early times \cite{Ito:2013ywa}
and to turn into a power law at later times \cite{Ito:2015mxa}
based on results obtained by Monte Carlo simulation of simplified models.
Classical solutions, which may describe the expanding behavior at
even later times, are also discussed \cite{Kim:2011ts,Kim:2012mw}.
These results are encouraging since they seem to suggest
that the Lorentzian version of the type IIB matrix model 
correctly describes the history of our Universe 
as it should for a nonperturbative
formulation of superstring theory.
See refs.~\cite{
Chaney:2015ktw,Chaney:2015mfa,Stern:2014aqa,%
Steinacker:2016vgf,Steinacker:2010rh,Klammer:2009ku,%
Yang:2015vna,Yang:2008fb,%
Tomita:2015let,Chatzistavrakidis:2014vsa}
for closely related work in this direction.

In this paper we investigate the effects of the infrared (IR) cutoffs,
which are inevitably introduced in the Lorentzian type IIB matrix model.
For that purpose, we generalize the form of the IR cutoffs as
$\tr \{ (A_0)^2 \}^p$ and $ \tr Q^p$ with
$Q =  \sum_{i=1}^{9} (A_i)^2 $,
where $p$ is a real positive parameter.
The previous choice corresponds to the $p=1$ case,
and in the $p\rightarrow \infty$ limit, the cutoffs 
constrain only the largest eigenvalues of $(A_0)^2$ and $Q$.
We first perform Monte Carlo simulation of a simplified model
for $1.0 \le p \le 1.5$,
and find that the results become 
universal
for $p > p_{\rm cr}$, where $p_{\rm cr} = 1.2 \sim 1.3$.
The previous results 
with $p=1$
agree with this universal behavior qualitatively but not quantitatively.
Some preliminary results have been reported 
in our proceedings article \cite{Ito:2015mem}.

The universality observed here suggests that the effects of the 
IR cutoffs disappear in the infinite-volume limit
for $p > p_{\rm cr}$.
In order to 
clarify this possibility,
we consider the Schwinger-Dyson equations (SDE)
and calculate each term by Monte Carlo simulation 
of the simplified model for $p=0.5$, 1.0 and 1.5.
We find for $p=1.5$ that the terms arising from the IR cutoffs
indeed decrease in magnitude compared with the other terms
in the SDE as the volume is increased.
This is not the case for $p=0.5$ and 1.0.

On the other hand, when $p$ becomes as large as $2.0$, 
we find that large-$N$ scaling behaviors do not show up,
which implies that we cannot take a sensible large-$N$ limit
unlike the cases with $p\le 1.5$.
This has something to do with the fact
that 
the number of eigenvalues $\alpha_i$ of $A_0$ 
that correspond to the time region in which the 
spontaneous symmetry breaking
occurs 
does not increase with $N$ for $p=2.0$.
Interestingly, for $p=2.0$, we observe a (5+1)d structure
instead of a (3+1)d structure observed for $p \le 1.5$.

Thus we conclude that there is a certain range of $p$ in which
a universal large-$N$ limit can be taken.
Within this range of $p$, the dynamical space-time dimensionality 
turns out to be
$(3+1)$,
at least in the simplified model.

The rest of this paper is organized as follows. 
In section \ref{sec:review} we introduce
the Lorentzian type IIB matrix model
with a generalized form of the IR cutoffs including 
the arbitrary parameter $p$.
In section \ref{sec:The-infrared-cutoff}
we explain the simplification of the model we adopt,
and present the results of Monte Carlo simulation
for $p \le 1.5$.
In particular, we show that a universal behavior is obtained
when $p$ is larger than some critical value.
In section \ref{sec:SDeq} we investigate
the IR cutoff effects by calculating
each term in the SDE by Monte Carlo simulation.
In section \ref{sec:larger-p} we show that large-$N$ scaling
behaviors do not show up for $p$ as large as 2.0.
Section \ref{sec:summary} is devoted
to a summary and discussions.

\section{The Lorentzian type IIB matrix model 
with a generalized IR cutoffs}
\label{sec:review}

The action of the type IIB matrix model 
is given by \cite{Ishibashi:1996xs}
\begin{eqnarray}
S & = & S_{\mathrm{b}}+S_{\mathrm{f}}\ ,\label{eq:S_likkt}\\
S_{\mathrm{b}} & = 
& -\frac{1}{4g^{2}}
\text{Tr}\left(\left[A_{\mu},A_{\nu}\right]
\left[A^{\mu},A^{\nu}\right]\right)\ ,
\label{eq:Sb}\\
S_{\mathrm{f}} & = & -\frac{1}{2g^{2}}
\text{Tr}\left(\Psi_{\alpha}
\left(\mathcal{C}\Gamma^{\mu}\right)_{\alpha\beta}
\left[A_{\mu},\Psi_{\beta}\right]\right)\ ,
\label{eq:Sf-1}
\end{eqnarray}
where 
$A_{\mu}$ $\left(\mu=0,\ldots,9\right)$
and 
$\Psi_{\alpha}$ $\left(\alpha=1,\ldots,16\right)$
are bosonic and fermionic $N\times N$ matrices, respectively,
both of which are traceless and Hermitian. 
The indices $\mu$ and $\nu$ are contracted
using the Lorentzian metric 
$\eta_{\mu\nu}=\text{diag}\left(-1,1,\ldots,1\right)$,
whereas the $16 \times 16$ matrices
$\Gamma^{\mu}$ and $\mathcal{C}$
are gamma-matrices 
and the charge conjugation matrix, respectively, 
after the Weyl projection in (9+1)-dimensions.
The model has manifest (9+1)-dimensional Lorentz symmetry,
where $A_{\mu}$ and $\Psi_{\alpha}$ transform as a vector and 
a Majorana-Weyl spinor, respectively.
The ``coupling constant'' $g$ is merely a scale parameter
since it can be absorbed by an appropriate rescaling 
of $A_{\mu}$ and $\Psi_\alpha$.
The Euclidean version can be obtained by making a ``Wick rotation''
$A_{0}=iA_{10}$, where $A_{10}$ is supposed to be Hermitian.

The partition function for the Lorentzian version 
is proposed in ref.~\cite{Kim:2011cr}
as 
\begin{equation}
Z=\int dAd\Psi\, e^{iS}\label{Z-Likkt1}
\end{equation}
with the action \eqref{eq:S_likkt}. The ``$i$'' in front of the
action is motivated from the fact that the string world-sheet metric
should also have a Lorentzian signature. By integrating out the fermionic
matrices, we obtain the Pfaffian 
\begin{equation}
\int d\Psi\, e^{iS_{\mathrm{f}}}=\text{Pf}\,\mathcal{M}\left(A\right)\ ,
\label{def-Pfaffian}
\end{equation}
which is real unlike in the Euclidean case \cite{Anagnostopoulos:2013xga}.
Note also that the bosonic action \eqref{eq:Sb} can be written as
\begin{equation}
S_{\mathrm{b}}=\frac{1}{4g^{2}}\text{Tr}\left(F_{\mu\nu}F^{\mu\nu}\right)
=\frac{1}{4g^{2}}\left\{ -2\text{Tr}\left(F_{0i}\right)^{2}+\text{Tr}\left(F_{ij}\right)^{2}\right\} \ ,
\label{decomp-Sb}
\end{equation}
where we have introduced the Hermitian matrices 
$F_{\mu\nu}=i\left[A_{\mu},A_{\nu}\right]$.
Since the two terms in the last expression of eq.~(\ref{decomp-Sb})
have opposite signs, $S_{\mathrm{b}}$ is
not positive semi-definite,
which makes the partition function \eqref{Z-Likkt1} divergent.
Let us recall that in the Euclidean case, $S_{\mathrm{b}}$ is
positive semi-definite, and the partition function is 
finite \cite{Krauth:1998xh,Austing:2001pk}.

In order to make the partition function \eqref{Z-Likkt1} finite,
we introduce IR cutoffs in both the temporal and spatial
directions as\footnote{One might be tempted to introduce
a Lorentz invariant IR cutoff 
of the form
$\frac{1}{N}\text{Tr}\left(A_{\mu}A^{\mu}\right)^p \leq L^{2p}$.
This does not work, however, because 
$A_{\mu} A^{\mu} = - (A_0)^2 + (A_i)^2$ can be small in magnitude
due to cancellations
between the two terms.}
\begin{alignat}{3}
\frac{1}{N}\text{Tr} \,  \{ \left(A_{0}\right)^{2}  \}^p
& \leq  \kappa^p \frac{1}{N}\text{Tr} \, \{ \left(A_{i}\right)^{2} \}^p  \ ,
\label{eq:t_cutoff}\\
\frac{1}{N}\text{Tr} \, \{ \left(A_{i}\right)^{2} \}^p
& \leq  L^{2p}\ ,
\label{eq:s_cutoff}
\end{alignat}
which generalizes the original one adopted in ref.~\cite{Kim:2011cr}
corresponding to the $p=1$ case.
In what follows, we set $L=1$ without loss of generality.
After some manipulation and rescaling of $A_{\mu}$, we can
rewrite the partition function \eqref{Z-Likkt1} as \cite{Kim:2011cr}
(See appendix A of ref.~\cite{Ito:2013ywa} for a refined argument.)
\begin{alignat}{3}
Z & =  \int dA\,\text{Pf}\mathcal{M}\left(A\right)
\delta\left(\frac{1}{N}\text{Tr}\left(F_{\mu\nu}F^{\mu\nu}\right)\right)
\delta\left(\frac{1}{N}\text{Tr}
\{ \left( A_{i}\right)^{2}\}^p-1\right)
\theta\left(\kappa^p -\frac{1}{N}\text{Tr}
\{  \left( A_{0}\right)^{2} \} ^p \right) \ ,
 \label{Z-Likkt2}
\end{alignat}
where $\theta\left(x\right)$ is the Heaviside step function. This
form allows us to perform Monte Carlo simulation
without the sign problem unlike in the Euclidean model.%
\footnote{Strictly speaking, 
the Pfaffian ${\rm Pf}\mathcal{M}$ in 
(\ref{Z-Likkt2})
can change its sign, but configurations with positive
Pfaffian dominate at large $N$.%
}

A peculiar feature of the Lorentzian version of the type IIB matrix model
is that one can extract the ``real-time dynamics'' by identifying
the eigenvalues of the temporal matrix $A_{0}$ as 
representing the time \cite{Kim:2011cr}.
For that purpose we use the SU($N$) symmetry of the
model to diagonalize the temporal matrix $A_{0}$ as 
\begin{equation}
A_{0}=\text{diag}\left(\alpha_{1},\ldots,\alpha_{N}\right)\ ,
\quad\quad \text{where} \, \alpha_{1}<\cdots<\alpha_{N}\ .
\label{eq:diagonal gauge}
\end{equation}
In this basis,
the spatial matrices $A_{i}$ generated by the Monte Carlo 
simulation of (\ref{Z-Likkt2})
turn out to have an approximate band-diagonal structure.
More precisely, there exists some integer
$n$ such that the elements of spatial matrices $\left(A_{i}\right)_{ab}$
for $\left|a-b\right|\geq n$ are much smaller than 
those for $\left|a-b\right|<n$.
Based on this observation, we may naturally consider $n\times n$
matrices 
\begin{equation}
\left(\bar{A}_{i}\right)_{IJ}\left(t \right)\equiv
\left(A_{i}\right)_{\nu+I,\nu+J}
\label{eq:def_abar}
\end{equation}
as representing the state of the 9d space at time $t$ defined by
\begin{equation}
t=\frac{1}{n}\sum_{I=1}^{n}\alpha_{\nu+I} \ ,
\label{eq:def_t}
\end{equation}
where $I,J=1,\ldots,n$ and $\nu=0,1,\ldots,N-n$.
For example,
we can define the extent of space at time $t$ as 
\begin{equation}
R^{2}\left(t\right)=
\left\langle \frac{1}{n}
\text{tr}\sum_{i}\left(\bar{A}_{i}\left(t\right)\right)^{2}\right\rangle \ ,
\label{eq:def_rsq}
\end{equation}
where the symbol $\text{tr}$ represents a trace over the $n\times n$
block. We also define the ``moment of inertia tensor'' 
\begin{equation}
T_{ij}\left(t\right)=
\frac{1}{n}\text{tr}\left(\bar{A}_{i}\left(t\right)\bar{A}_{j}
\left(t\right)\right)\ ,
\label{eq:def_tij}
\end{equation}
which is a $9\times9$ real symmetric matrix. 
The eigenvalues of $T_{ij}\left(t\right)$,
which we denote by $\lambda_{i}\left(t\right)$ with the order 
\begin{equation}
\lambda_{1}\left(t\right)>\lambda_{2}\left(t\right)>\cdots>\lambda_{9}\left(t\right)\ ,
\end{equation}
represent the spatial extent in each of the nine directions at time
$t$. 
The block size $n$ used in calculating
quantities such as (\ref{eq:def_rsq}) and (\ref{eq:def_tij})
by Monte Carlo simulation
is determined as described in section 5 of ref.~\cite{Ito:2015mxa}.

In actual simulation, it is convenient to ``gauge fix'' 
the SU($N$) symmetry by the condition (\ref{eq:diagonal gauge}).
The usual Fadeev-Popov procedure for the gauge fixing implies that
the integration $\int dA_{0}$ in \eqref{Z-Likkt2} should be replaced
by $\int \prod_{k=1}^{N}d\alpha_{k}\,\Delta(\alpha)^{2}$, where 
\begin{alignat}{3}
\Delta(\alpha)
\equiv\prod_{a>b}^{N}\left(\alpha_{a}-\alpha_{b}\right)
\label{def-vdm}
\end{alignat}
is the van der Monde determinant.
The delta functions and the step function in \eqref{Z-Likkt2}
are replaced by the Gaussian-type potentials in the action given by
\begin{eqnarray}
S^{(C)} 
& = & \frac{1}{2}\gamma^{(C)}N^{2}
\left(\frac{1}{N}\text{Tr} (F_{\mu\nu}F^{\mu\nu}) \right)^{2}\;,
\label{eq:strf}\\
S^{(L)} 
& = & \frac{1}{2}\gamma^{(L)}N^{2}
\left(\frac{1}{N}\text{Tr}
\left[\{ \left(A_{i}\right)^{2}\}^{p}\right]-1\right)^{2}\;,
\label{eq:strai}\\
S^{(\kappa)} & = & 
\begin{cases}
\frac{1}{2}\gamma^{(\kappa)}N^{2}
\left(\frac{1}{N}\text{Tr}
\left[ \{ \left(A_{0}\right)^{2} \}^{p}\right]-\kappa^{p}\right)^{2}
& \text{for }\frac{1}{N}
\text{Tr}\left[\{ \left(A_{0}\right)^{2}\}^{p} \right]>\kappa^{p} \ ,\\
0 & \text{otherwise} \ ,
\end{cases}
\label{eq:stra0-1}
\end{eqnarray}
where the coefficients $\gamma^{(C)},\,\gamma^{(L)}$ and $\gamma^{(\kappa)}$
are taken to be large enough to make the generated configurations
satisfy the constraints with good accuracy.
Thus, the partition function \eqref{Z-Likkt2} is replaced 
by\footnote{As a yet another technical detail, 
we use a potential for stabilizing the peak of $R^2(t)$
defined by (\ref{eq:def_rsq}), whose position would otherwise fluctuate 
slowly as the simulation proceeds.
See appendix B of ref.~\cite{Ito:2013ywa} for the details.
This is done just for the sake of effective measurements,
and it does not affect the properties of the model.
For instance, we have explicitly checked that a term that appears from
this peak-stabilizing potential is negligible compared with the other terms
in the SDE investigated in Section \ref{sec:SDeq}.
}
\begin{equation}
Z=\int\prod_{i=1}^{9}dA_{i}
\prod_{k=1}^{N}d\alpha_{k}\,\Delta(\alpha)^{2} \, 
\text{Pf}\,\mathcal{M}\left(A\right)
e^{-(S^{(C)}+S^{(L)}+S^{(\kappa)})} \ .
\label{Z-Likkt3}
\end{equation}
This model can be investigated by Monte Carlo simulation
as described in appendix B of ref.~\cite{Ito:2013ywa} for $p=1$.


\begin{figure}[t]
\centering{}
\includegraphics[width=7cm]{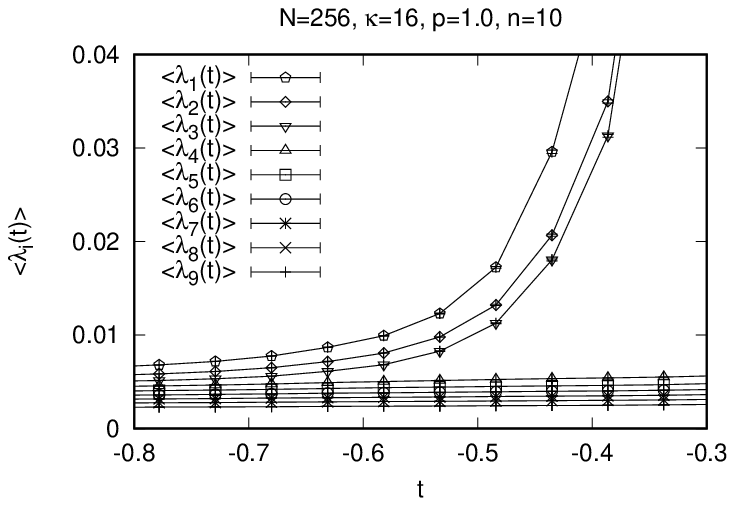}
\includegraphics[width=7cm]{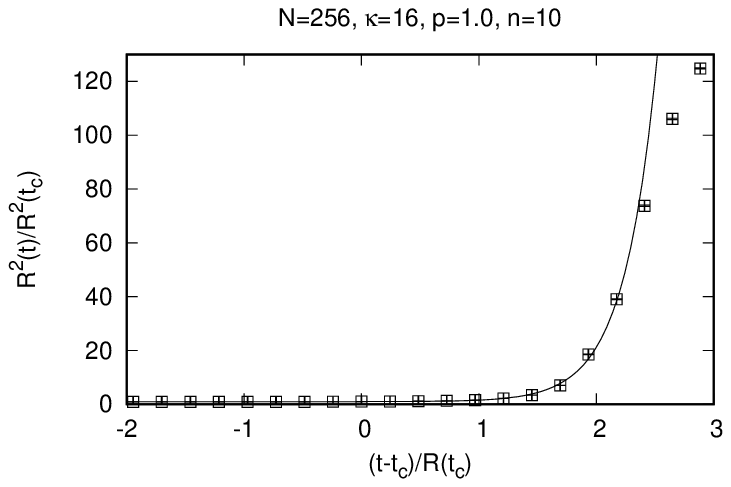}
\caption{
(Left) 
The expectation values of 
the eigenvalues $\lambda_{i}\left(t\right)$
of $T_{ij}\left(t\right)$ are plotted against $t$ 
for $p=1$ with $N=256$ and $\kappa=16$, where we use the
block size $n=10$. 
The lines are drawn to guide the eye.
(Right) The extent of space 
$R^{2}\left(t\right)$
normalized by $R^{2}\left(t_{\mathrm{c}}\right)$ is plotted against
$x=\left(t-t_{\mathrm{c}}\right)/R\left(t_{\mathrm{c}}\right)$
for the same set of parameters.
The solid line is a fit to the exponential behavior 
$f(x) = a + (1-a)\exp (bx)$ with a constraint $f(0)=1$,
where $a=0.983(3)$ and $b=3.56(11)$.
\label{fig:t_r_n256k16p1}}
\end{figure}

\section{Universality in the results for various $p$}
\label{sec:The-infrared-cutoff}

Let us investigate how the results of the model
(\ref{Z-Likkt3}) depend on the parameter $p$ introduced in
the cutoffs (\ref{eq:t_cutoff}) and (\ref{eq:s_cutoff}).
Here, we adopt a simplification \cite{Ito:2013ywa},
which amounts to replacing the Pfaffian in (\ref{Z-Likkt3}) as
\begin{equation}
\text{Pf}\mathcal{M}\left(A\right)
\Longrightarrow  \Delta^{16}\left(\alpha\right) \ ,
\label{eq:vdm}
\end{equation}
where $\Delta\left(\alpha\right)$ is 
the van der Monde determinant defined in (\ref{def-vdm}).
This simplification occurs when one omits the dependence of 
$\text{Pf}\mathcal{M}\left(A\right)$ on the spatial matrices $A_i$,
which makes sense at early times, where
the expansion of space has not proceeded much.
The partition function of the simplified model is given by
\begin{equation}
Z=\int\prod_{i=1}^{9}dA_{i}
\prod_{k=1}^{N}d\alpha_{k}\,\Delta(\alpha)^{18} \, 
e^{-(S^{(C)}+S^{(L)}+S^{(\kappa)})} \ ,
\label{Z-Likkt4}
\end{equation}
and the computational cost 
is
considerably reduced from that of the original model (\ref{Z-Likkt3}).


Let us first consider the case in which 
$p=1$ is used for the IR cutoffs
(\ref{eq:t_cutoff}) and (\ref{eq:s_cutoff})
as is done in all the previous work.
Figure~\ref{fig:t_r_n256k16p1} (Left)
shows the expectation values of
the eigenvalues $\lambda_{i}\left(t\right)$ of
$T_{ij}\left(t\right)$ 
defined by (\ref{eq:def_tij})
as a function of $t$ for the simplified model 
with $N=256$ and $\kappa=16$.
We find that the SO(9) symmetry is spontaneously
broken down to SO(3) at a critical time $t_{\mathrm{c}}=-0.68014(7)$. 
(Precise definition of $t_{\mathrm{c}}$
is given in section 3 of ref.~\cite{Ito:2013ywa}.)
Since the extent of space $R^{2}\left(t_{\mathrm{c}}\right)=0.04099(4)$
at the critical time $t_{\mathrm{c}}$
is a physical quantity which is dimensionful,
we use it to fix the scale of the system.
In Fig.~\ref{fig:t_r_n256k16p1} (Right), we plot the extent
of space $R^{2}\left(t\right)$ normalized
by $R^{2}\left(t_{\mathrm{c}}\right)$ against 
$x=\left(t-t_{\mathrm{c}}\right)/R\left(t_{\mathrm{c}}\right)$
for the same $N$ and $\kappa$. 
The result can be nicely fitted to an exponential function.
(Note that the data points 
at late times are
affected by finite $N$ effects as one can see from the 
large-$N$ scaling behaviors in Fig.~\ref{fig:Rt_p05} below.)
Similar behaviors were observed previously
in the (5+1)d version
of the simplified model with the matrix size $N\le64$ \cite{Ito:2013ywa}.



\begin{figure}[H]
\begin{center}
\includegraphics[width=13cm]{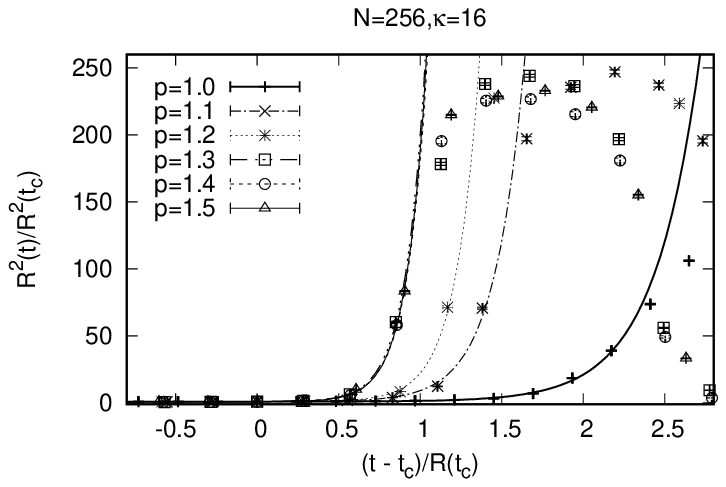}
\includegraphics[width=13cm]{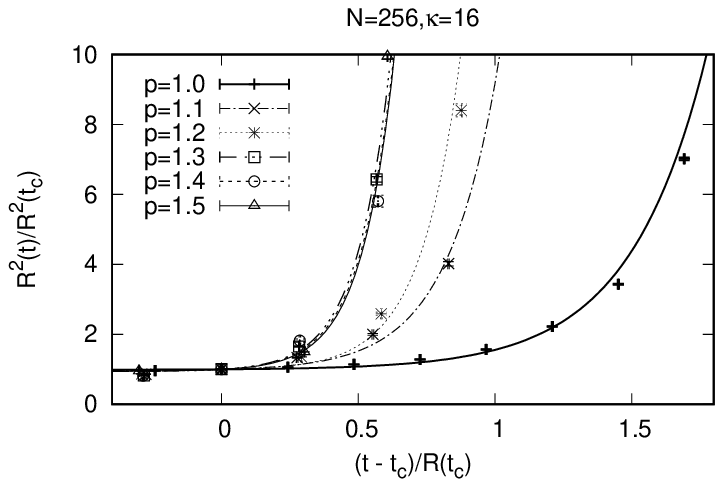}
\end{center}
\caption{(Top) The extent of space $R^{2}\left(t\right)$
normalized by $R^{2}\left(t_{\mathrm{c}}\right)$
is plotted against
$x=\left(t-t_{\mathrm{c}}\right)/R\left(t_{\mathrm{c}}\right)$
for $1.0\leq p\leq1.5$ with $N=256$ and $\kappa=16$.
The parameters used to make these plots
are given in table~\ref{tab:various_p}.
The lines are fits to
$R^{2}\left(t\right)/R^{2}
\left(t_{\mathrm{c}}\right)=a+\left(1-a\right)\exp\left(bx\right)$.
The values of the fitting parameters
$a$ and $b$ obtained by the fits
are also presented in table~\ref{tab:various_p}.
(Bottom) Zoom up of the plot at the top.
\label{fig:rt_various_p}}
\end{figure}

Next we show our results for $p\ne 1$.
In Fig.~\ref{fig:rt_various_p} we plot the extent of space 
$R^{2}\left(t\right)$ normalized by
$R^{2}\left(t_{\mathrm{c}}\right)$
against $x=\left(t-t_{\mathrm{c}}\right)/R\left(t_{\mathrm{c}}\right)$
for various values of $p$ within $1\leq p\leq1.5$.
While the results exhibit certain $p$-dependence,
qualitative behaviors such as the exponential expansion 
remain the same as those for $p=1.0$.
We have also confirmed that 
only three directions start to expand
at some critical time for the values of $p$ within this region. 
What is most remarkable in these plots is that
the data points for $\mbox{~}1.3\leq p\leq1.5\mbox{~~}$
lie on a single
\noindent curve except for the region of $t$ in which
$R^2(t)$ approaches its maximum.
This universality suggests that the IR cutoffs are not affecting
the results
for these values of $p$
except near the spatial ``boundary'', where the cutoff effects 
should, of course, be visible.

\begin{table}[t]
\centering{}%
\begin{tabular}{|c|c|c||c|c|c||c|c|}
\hline 
$p$ & $N$ & $\kappa$ & $n$ & $t_{\mathrm{c}}$ & 
$R^{2}\left(t_{\mathrm{c}}\right)$ & $a$ & $b$\tabularnewline
\hline 
\hline 
1.0 & 256 & 16 & 10 & -0.68014(7) & 0.04099(04) & 0.983(03) & 3.56(11)\tabularnewline
\hline 
1.1 & 256 & 16 & 6 & -0.39307(6) & 0.03213(14) & 0.961(18) & 5.36(39)\tabularnewline
\hline 
1.2 & 256 & 16 & 6 & -0.34441(6) & 0.02904(16) & 0.976(12) & 6.82(53)\tabularnewline
\hline 
1.3 & 256 & 16 & 6 & -0.29213(8) & 0.03055(11) & 0.940(12) & 8.10(28)\tabularnewline
\hline 
1.4 & 256 & 16 & 6 & -0.23933(8) & 0.02940(19) & 0.944(27) & 8.07(63)\tabularnewline
\hline 
1.5 & 256 & 16 & 6 & -0.23593(7) & 0.02579(02) & 0.950(11) & 8.24(30)\tabularnewline
\hline 
\end{tabular}
\caption{The block size $n$, the critical time $t_{\mathrm{c}}$ 
and the extent of space 
$R^{2}\left(t_{\mathrm{c}}\right)$ at the critical time, which
are used to make the plots in Fig.~\ref{fig:rt_various_p},
are given for each $p$.
We also present the values of $a$ and $b$
obtained by fitting 
$R^{2}\left(t\right)/R^{2}\left(t_{\mathrm{c}}\right)$
to $f\left(x\right)=a+\left(1-a\right)\exp\left(bx\right)$ 
with $x=\left(t-t_{\mathrm{c}}\right)/R\left(t_{\mathrm{c}}\right)$
for each $p$.
\label{tab:various_p}}
\end{table}

\begin{figure}[t]
\begin{centering}
\includegraphics[width=7cm]{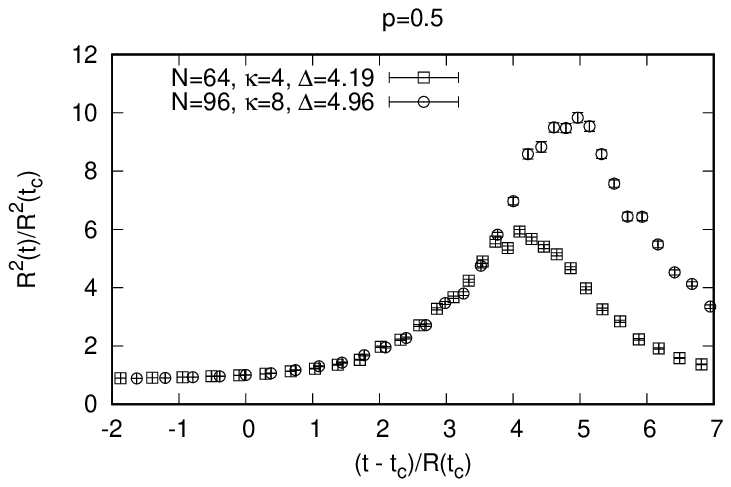}
\includegraphics[width=7cm]{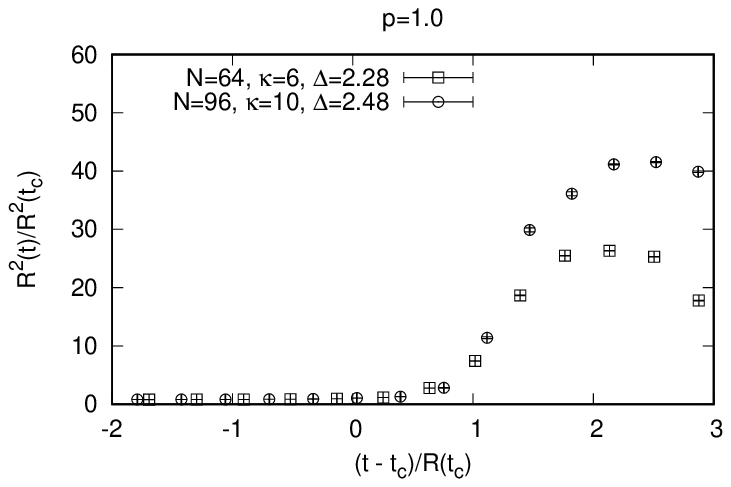}
\par\end{centering}

\centering{}
\includegraphics{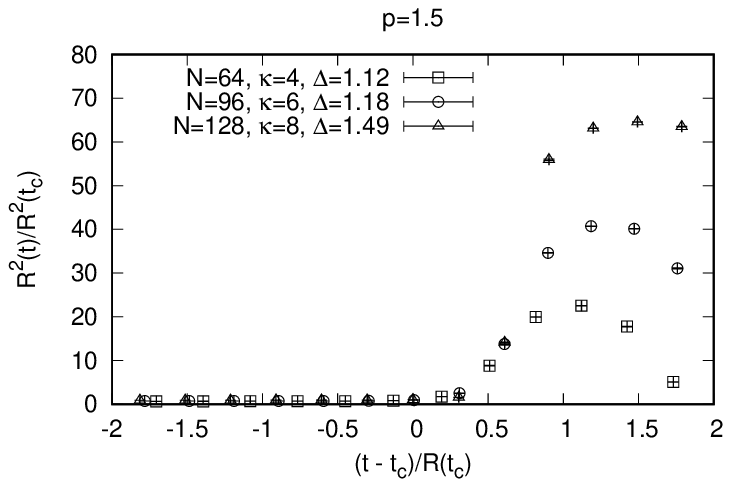}
\caption{The extent of 
space $R^{2}\left(t\right)$ normalized by $R^{2}\left(t_{\mathrm{c}}\right)$
is plotted against $x=\left(t-t_{\text{c}}\right)/R\left(t_{\mathrm{c}}\right)$
for $p=0.5$ (Top-Left), $p=1.0$ (Top-Right) and $p=1.5$ (Bottom).
The parameters $N$ and $\kappa$ are chosen as
in table~\ref{tab:infinite_volume} for each $p$ so that 
the lattice spacing $\varepsilon$ is kept almost constant in $N$, while
the volume $\Delta$ in the temporal direction increases with $N$.
\label{fig:Rt_p05}}
\end{figure}

In the next section, we examine this interpretation directly
by investigating the IR cutoff effects
through the SDE.
In particular, we show that the IR cutoff effects actually 
decrease in magnitude for large enough $p$
as we take the infinite-volume limit.
In the remainder of this section, we discuss how we take this limit.
First we define
the ``volume'' $\Delta$
and the ``lattice spacing''
$\varepsilon$
in the temporal direction by 
\begin{equation}
\Delta\equiv
\frac{t_{\text{peak}}-t_{\mathrm{c}}}{R\left(t_{{\rm c}}\right)} \ ,
\quad  \quad
\varepsilon=\frac{\Delta}{\nu} \ ,
\label{def-Delta}
\end{equation}
where $t_{\rm peak}$ represents the time $t$ at which $R^2(t)$ becomes 
maximum, and 
$\nu$ is the number of data points within $\Delta$. 
The infinite-volume limit corresponds to increasing 
$\Delta$ with fixed $\varepsilon$, 
while the continuum limit corresponds to
decreasing $\varepsilon$ with fixed $\Delta$.
By tuning the cutoff parameter $\kappa$ as one increases $N$, 
one can take these limits
separately or simultaneously.

When we investigate the SDE in section \ref{sec:SDeq},
we need to take the infinite-volume limit since 
the IR cutoff effects are expected to disappear in that limit. 
For that purpose, we tune the cutoff parameter $\kappa$ 
as we increase $N$
for each $p$
so that the lattice spacing is kept almost constant in $N$.
In table \ref{tab:infinite_volume}, we show the values of $\kappa$
thus obtained together with the lattice spacing $\varepsilon$ and
the volume $\Delta$ measured using (\ref{def-Delta}).
In Fig.~\ref{fig:Rt_p05}, we plot
the extent of space $R^{2}\left(t\right)$ 
normalized by $R^{2}\left(t_{\mathrm{c}}\right)$
against $x=\left(t-t_{\text{c}}\right)/R\left(t_{\mathrm{c}}\right)$
for the set of parameters given in table \ref{tab:infinite_volume}.
We find that the horizontal distance of the data points is almost 
independent of $N$, while the extent in the temporal and spatial 
directions grows with $N$. 

\begin{figure}[t]
\begin{centering}
\includegraphics[width=7cm]{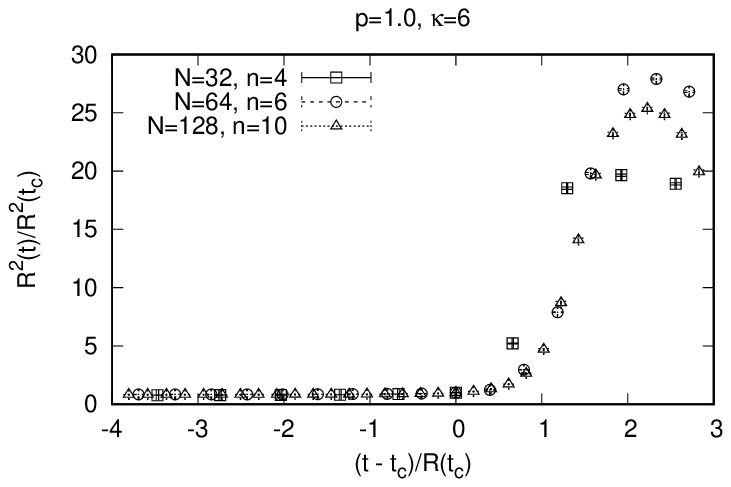}
\includegraphics[width=7cm]{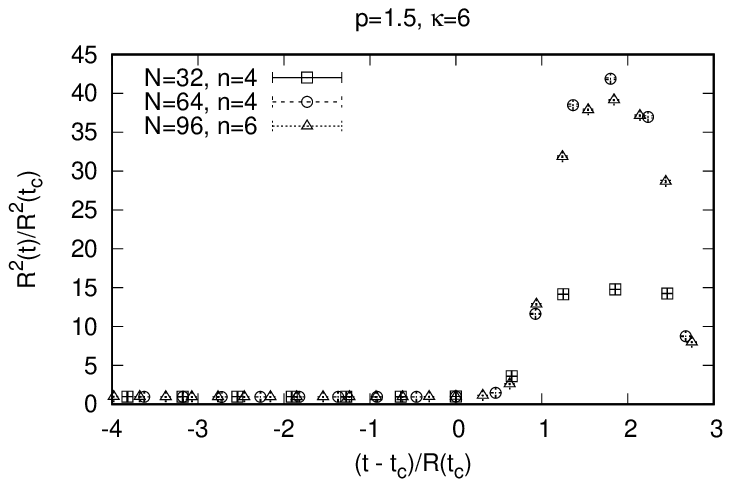}
\par\end{centering}
\centering{}
\includegraphics{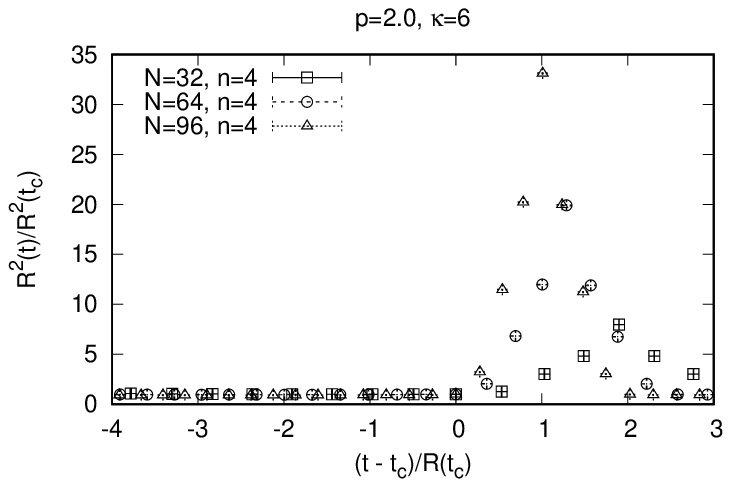}
\caption{
The extent of space 
$R^{2}\left(t\right)$
normalized by $R^{2}\left(t_{\mathrm{c}}\right)$ is plotted against
$x=\left(t-t_{\mathrm{c}}\right)/R\left(t_{\mathrm{c}}\right)$
for $\kappa=6$ with $p=1.0$ (Top-Left), $p=1.5$ (Top-Right)
and $p=2.0$ (Bottom).
\label{fig:comp-p-kappa6}}
\end{figure}

Note also that Fig.~\ref{fig:Rt_p05} 
exhibits
a scaling region in which
the data for different $N$ lie on top of each other.
In making these plots, we have shifted the value of $t_{\rm c}$
slightly so that the observed scaling behavior is optimized,
which is legitimate taking into account the ambiguity 
in the definition of $t_{\rm c}$ at finite $N$. 
Similar shifts are used also in 
Fig.~\ref{fig:infinite_volume_tmp}, 
Fig.~\ref{fig:comp-p-kappa6} (Top-Left),
Fig.~\ref{fig:comp-p-kappa6} (Top-Right) and
Fig.~\ref{fig:infinite_volume_spc}, where we 
discuss large-$N$ asymptotic behaviors.

\begin{table}[H]
\begin{centering}
\begin{tabular}{|c|c|c||c|c|c||c|c|}
\hline 
$p$ & $N$ & $\kappa$ & $n$ & $t_{\mathrm{c}}$ & 
$R^{2}\left(t_{\mathrm{c}}\right)$ & $\varepsilon$ & $\Delta$\tabularnewline
\hline 
\hline 
0.5 & 64 & 4 & 
8 & -1.27274(45) & 0.15612(90) &
0.28 & 4.19\tabularnewline
\hline 
0.5 & 96 & 8 & 
8 & -1.54629(53) & 0.09735(66) &
0.27 & 4.96\tabularnewline
\hline 
\hline 
1.0 & 64 & 6 & 
6 & -0.71603(18) & 0.09398(16) &
0.38 & 2.28\tabularnewline
\hline 
1.0 & 96 & 10 &
8 & -0.83132(75) & 0.07762(39) &
 0.35 & 2.48\tabularnewline
\hline 
\hline 
1.5 & 64 & 4 & 
4 & -0.38586(16) & 0.07797(18) &
0.312 & 1.12\tabularnewline
\hline 
1.5 & 96 & 6 & 
5 & -0.35286(14) & 0.06658(26) &
0.294 & 1.18\tabularnewline
\hline 
1.5 & 128 & 8 & 
6 & -0.33829(19) & 0.05349(14) &
0.298 & 1.49\tabularnewline
\hline 
\end{tabular}
\par\end{centering}

\caption{The set of parameters $N$ and $\kappa$
chosen for each $p$
in such a way that
the lattice spacing $\varepsilon$
is kept almost constant in $N$, while the volume $\Delta$ 
in the temporal direction increases with $N$.
The block size $n$, the critical time $t_{\mathrm{c}}$ 
and the extent of space 
$R^{2}\left(t_{\mathrm{c}}\right)$ at the critical time
are also given.
\label{tab:infinite_volume}}
\end{table}

\section{Probing the IR cutoff effects by the SDE}
\label{sec:SDeq}



As we mentioned above,
the universal behavior observed in Fig.~\ref{fig:rt_various_p}
suggests that 
the effects of the IR cutoffs
(\ref{eq:t_cutoff}) and (\ref{eq:s_cutoff})
vanish in the infinite-volume limit for sufficiently large $p$.
In order to clarify this possibility,
we investigate
the effects directly by using the SDE.

Here we rewrite the partition function (\ref{Z-Likkt4}) as
\begin{align}
Z & =
\int dA \, d\alpha \,
e^{-S} \ , 
\label{eq:partition_func-2}
\\
S & = S^{(C)}+S^{(L)}+S^{(\kappa)} + S^{(\alpha)} \ ,
\label{eq:decompose_action}
\end{align}
where $S^{(C)}$, $S^{(L)}$ and $S^{(\kappa)}$ are defined by
(\ref{eq:strf}), (\ref{eq:strai}) and (\ref{eq:stra0-1}), respectively,
and $S^{(\alpha)}$ is defined by
\begin{align}
S^{(\alpha)} &= 
-18\sum_{a>b}\ln\left(\alpha_{a}-\alpha_{b}\right)  \ .
\label{eq:svdm}
\end{align}
Let us then consider the SDE 
\begin{align}
\frac{1}{Z}\int dA \, d\alpha \,
\frac{\partial}{\partial \alpha_b} 
\left( \alpha_a e^{-S} \right) & =0 \ ,
\\
\frac{1}{Z}\int dA \, d\alpha \,
\frac{\partial}{\partial (A_j)_{cd}} 
\left( (A_i)_{ab}  e^{-S} \right) & =0 \ . 
\label{sdeq}
\end{align}
Contracting some indices, we obtain the identities
\begin{align}
\left\langle \alpha_{a}
\frac{\partial S}{\partial\alpha_{b}}\right\rangle 
&=\delta_{ab}-\frac{1}{N} \ ,
\label{sdeq_temp-2}
\\
\frac{1}{9N}
\sum_{i=1}^{9} \sum_{c=1}^{N}
 \left\langle \left(A_{i}\right)_{ac}
\frac{\partial S}{\partial\left(A_{i}\right)_{bc}}\right\rangle 
&=\left(1-\frac{1}{N^{2}}\right)\delta_{ab}  \ ,
\label{sdeq_spc-2}
\end{align}
which should be satisfied for each $a$ and $b$.
Below we focus on the identities corresponding to the $a=b$ case.
Corresponding to the decomposition (\ref{eq:decompose_action})
of the action, we obtain
\begin{eqnarray}
G^{(C)}_a + G^{(\kappa)}_a + G^{(\alpha)}_a & = & 1-\frac{1}{N} \ ,
\label{eq:sdeq_temp-3} \\
H^{(C)}_a +H^{(L)}_a  & = & 1-\frac{1}{N^{2}}\ ,
\label{eq:sdeq_spc-3}
\end{eqnarray}
where $a=1 , \cdots , N$ and we have defined
\begin{align}
& G^{(C)}_a \equiv 
\left\langle \alpha_{a}\frac{\partial S^{(C)}}{\partial\alpha_{a}}
\right\rangle  \ , \quad
G^{(\kappa)}_a \equiv
\left\langle \alpha_{a}
\frac{\partial S^{(\kappa)}}{\partial\alpha_{a}}\right\rangle 
\ , \quad
G^{(\alpha)}_a \equiv
\left\langle \alpha_{a}
\frac{\partial S^{(\alpha)}}{\partial\alpha_{a}}\right\rangle \ ,
\label{eq:eq:temp_cutoff_term}
\\
& H^{(C)}_{a} \equiv
\frac{1}{9N}\sum_{i=1}^{9}\sum_{c=1}^{N}
\left\langle \left(A_{i}\right)_{ac}
\frac{\partial S^{(C)}}{\partial\left(A_{i}\right)_{ba}}
\right\rangle 
\ ,
\quad
H^{(L)}_{a}
\equiv 
\frac{1}{9N}\sum_{i=1}^{9}\sum_{c=1}^{N}
\left\langle \left(A_{i}\right)_{ac}
\frac{\partial S^{(L)}}{\partial\left(A_{i}\right)_{ba}}\right\rangle 
\ .
\label{eq:spc_cutoff_term}
\end{align}



\begin{figure}[H]
\begin{centering}
\includegraphics{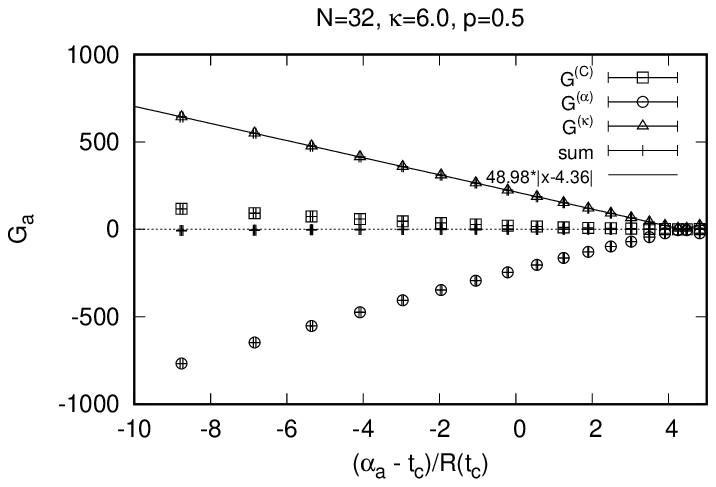}
\includegraphics{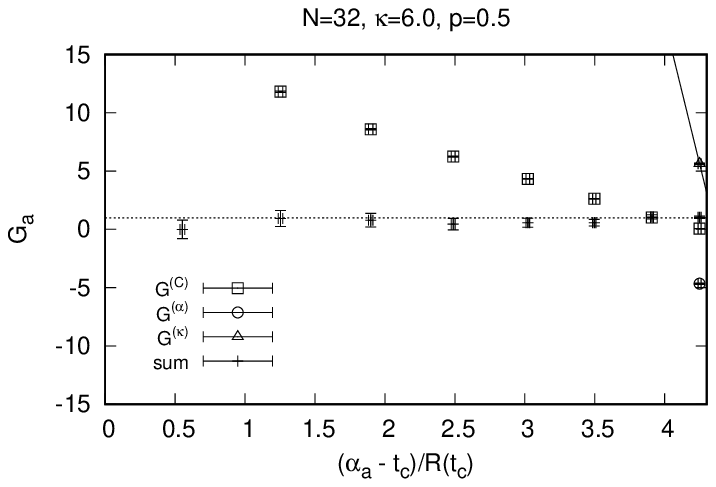}
\includegraphics{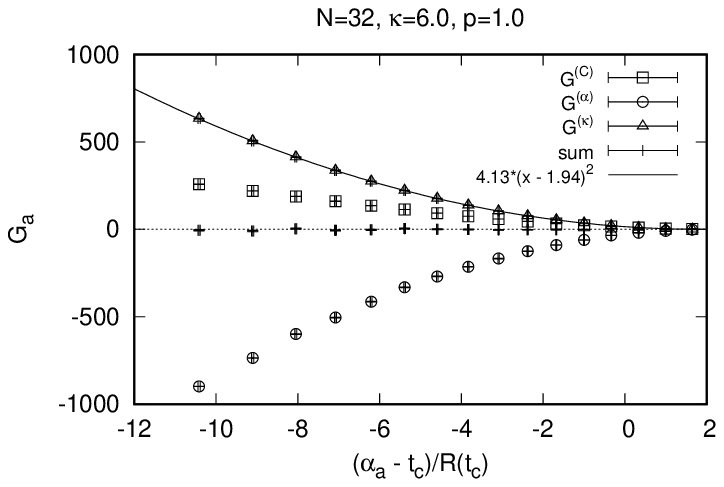}
\includegraphics{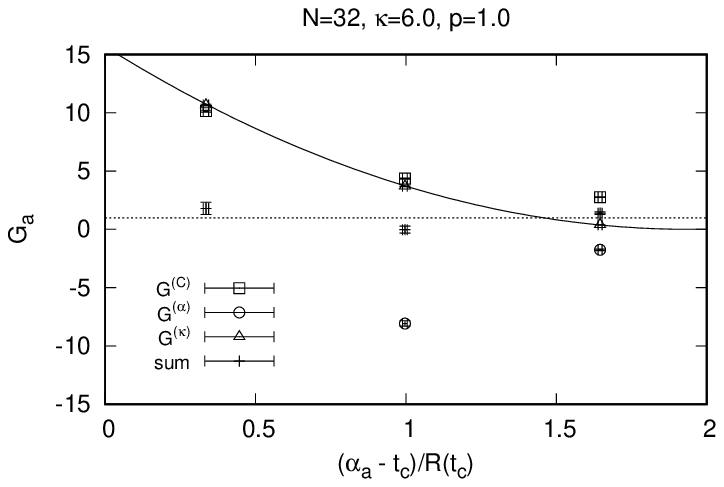}
\includegraphics{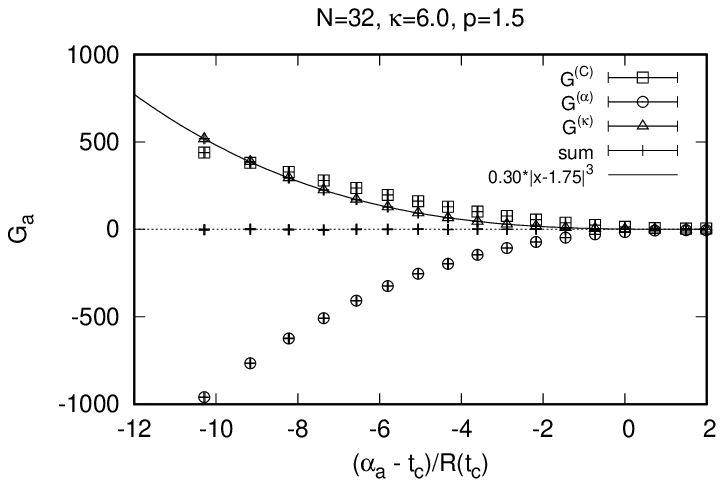}
\includegraphics{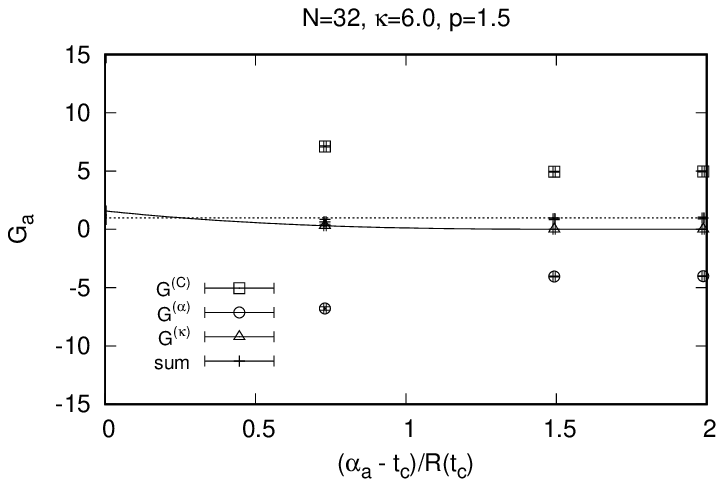}
\par\end{centering}
\caption{The terms
$G^{(C)}_a$, $G^{(\kappa)}_a$ and $G^{(\alpha)}_a$
in the temporal SDE
(\ref{eq:sdeq_temp-3}) are plotted against 
$x=\left(\alpha_{a}-t_{\mathrm{c}}\right)/R\left(t_{\mathrm{c}}\right)$
for $p=0.5$ (Top), $p=1.0$ (Middle) and $p=1.5$ (Bottom)
with $N=32$ and $\kappa=6$.
The plots on the right are zoom up of the plots on the left 
in the $t\ge 0$ region.
We also plot 
the sum of the three terms,
which agrees well with $1-1/N$ represented by the dotted lines.
The solid lines represent fits of $G^{(\alpha)}_a$
to the $(\alpha_{a})^{2p}$ behavior.
\label{fig:sdeq_check_temp}}
\end{figure}

In Fig.~\ref{fig:sdeq_check_temp}, 
we plot $G^{(C)}_a$, 
$G^{(\kappa)}_a$,
$G^{(\alpha)}_a$
and their sum against 
$\left(\alpha_{a}-t_{\mathrm{c}}\right)/R\left(t_{\mathrm{c}}\right)$
for $N=32$ and $\kappa=6$ with 
$p=0.5,\,1.0$ and $1.5$. 
We find that the sum of the three terms is constant and
agrees with $1 - \frac{1}{N}$,
which implies that the temporal SDE \eqref{eq:sdeq_temp-3} 
is satisfied for all $\alpha_a$. 
We also find that 
$G^{(\kappa)}_a$ can be nicely fitted to 
the $(\alpha_{a})^{2p}$ behavior.
This is understandable since
$G^{(\kappa)}_a$ 
can be written explicitly as
\begin{equation}
G^{(\kappa)}_a
=c\left(\left(\alpha_{a}\right)^{2p}
-\frac{\alpha_{a}}{N}\sum_{b=1}^N \alpha_{b}
\left(\alpha_{b}\right)^{2(p-1)}\right) \ ,
\label{eq:temp_cutoff_term_explicit}
\end{equation}
where the coefficient is given as
\begin{align}
c=2p\gamma^{(\kappa)}N\left[\frac{1}{N}\sum_{b=1}^{N}
\left(\alpha_{b}^{2}\right)^{p}- \kappa ^p \right] \ , 
\end{align}
and the first term in (\ref{eq:temp_cutoff_term_explicit}) actually 
dominates.

\begin{figure}[t]
\begin{centering}
\includegraphics[width=7cm]{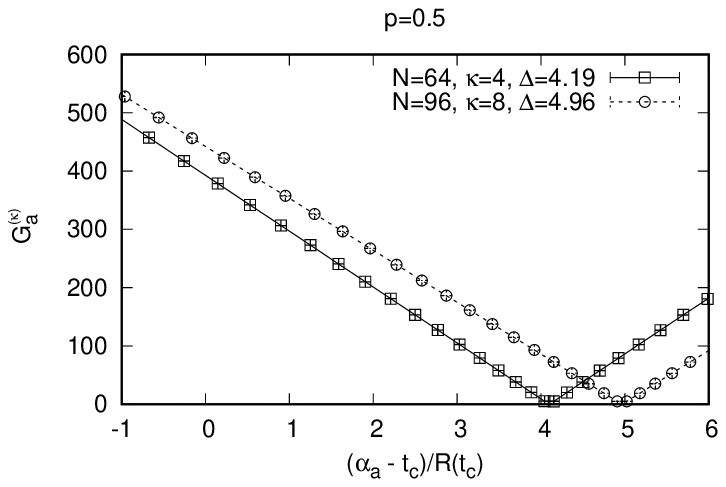}
\includegraphics[width=7cm]{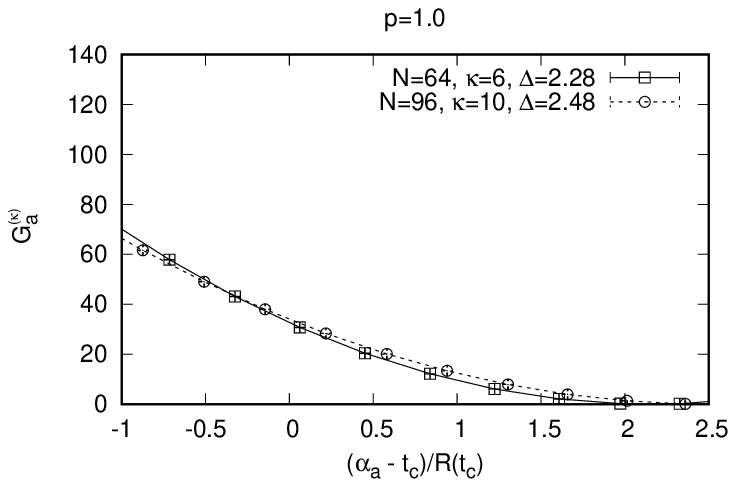}
\par\end{centering}

\centering{}
\includegraphics{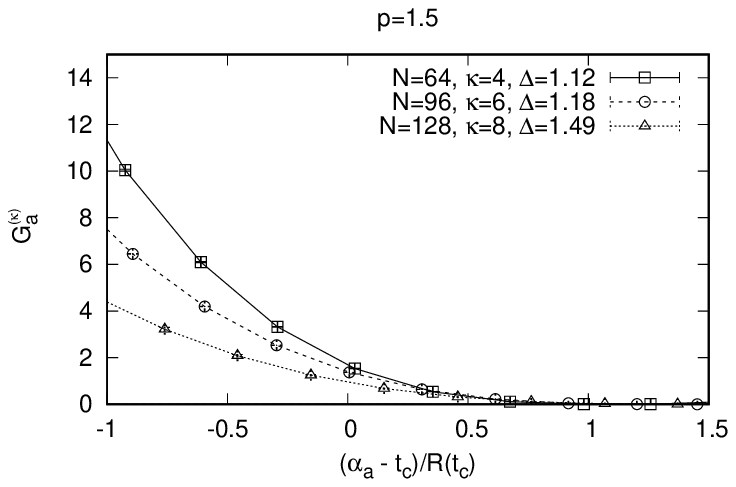}
\caption{
The IR cutoff term $G^{(\kappa)}_a$
in the temporal SDE (\ref{eq:sdeq_temp-3}) is plotted against
$x=\left(\alpha_{a}-t_{\mathrm{c}}\right)/R\left(t_{\mathrm{c}}\right)$
for $p=0.5$ (Top-Left), $p=1.0$ (Top-Right) and $p=1.5$ (Bottom)
with $\kappa$ and $N$
given in table \ref{tab:infinite_volume}.
The lines are drawn to guide the eye.
\label{fig:infinite_volume_tmp}}
\end{figure}

From this figure, we find that
the effects of the IR cutoffs in the temporal direction
represented by $G^{(\kappa)}_a$ become large towards the 
boundary in the temporal direction
represented by the left-most point in the plots on the left.
However, the IR cutoff effects are suppressed as one goes away
from the boundary, in particular for large $p$ as expected 
from (\ref{eq:temp_cutoff_term_explicit}).

Let us then consider the infinite-volume limit discussed at the
end of section \ref{sec:The-infrared-cutoff} and see
how the IR cutoff effects behave in that limit.
In Fig.~\ref{fig:infinite_volume_tmp}, we plot
$G^{(\kappa)}_a$
against
$x=\left(\alpha_{a}-t_{\mathrm{c}}\right)/R\left(t_{\mathrm{c}}\right)$
for $p=0.5$, $p=1.0$ and $p=1.5$.
The parameters $N$ and $\kappa$ are chosen
as in table \ref{tab:infinite_volume}
so that the lattice spacing in the temporal direction 
is kept almost constant in $N$,
while the volume $\Delta$ increases with $N$.
We find that $G^{(\kappa)}_a$
increases with the volume $\Delta$ for $p=0.5$,
whereas it decreases with the volume $\Delta$ for $p=1.5$.
For $p=1.0$, the results of $G^{(\kappa)}_a$
for different $\Delta$ lie almost on top of each other.

Our results for the SDE in the spatial direction
are presented in appendix \ref{sec:spatialSDE},
where we find that 
the term $H^{(L)}_a$ in (\ref{eq:sdeq_spc-3}),
which comes from the spatial cutoff, decreases in magnitude
for $p=1.5$ as the infinite-volume limit is taken.
This is not the case for $p=0.5$ and $p=1.0$.




\section{Absence of large-$N$ scaling behavior for $p=2.0$}
\label{sec:larger-p}

In this section we discuss the results obtained for larger $p$.
Here we focus on $p=2.0$, in which case
the IR cutoffs in (\ref{eq:t_cutoff}) and (\ref{eq:s_cutoff})
involve a term with the same canonical dimension 
as the bosonic action (\ref{eq:Sb}).
In Fig.~\ref{fig:comp-p-kappa6} we plot 
the extent of space 
$R^{2}\left(t\right)$ normalized by
$R^{2}\left(t_{\mathrm{c}}\right)$
for $\kappa=6$ with $p=1.0$ (Top-Left), $p=1.5$ (Top-Right)
and $p=2.0$ (Bottom).
While we observe large-$N$ scaling behaviors for 
$p=1.0$ and $p=1.5$ as we have already seen in Fig.~\ref{fig:Rt_p05},
this turns out to be not the case for $p=2.0$.
Hence, we cannot take a sensible large-$N$ limit for $p=2.0$.
In fact, the number of data points in the region where the 
spontaneous breaking of SO(9) symmetry occurs 
increases with $N$ for $p=1.0$ and $p=1.5$, but not for $p=2.0$,
where we have seven data points in the 
symmetry broken region for all $N$.
We consider that this is the reason why large-$N$ scaling behaviors
do not show up for $p=2.0$.

\begin{figure}[t]
\centering{}
\includegraphics[width=9cm]{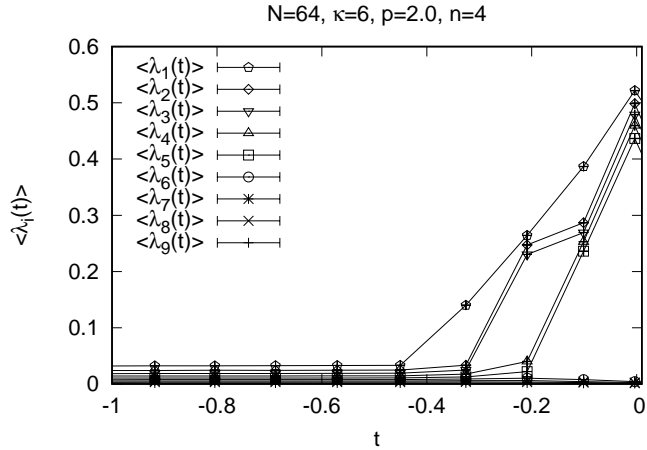}
\caption{
The expectation values of 
the eigenvalues $\lambda_{i}\left(t\right)$
of $T_{ij}\left(t\right)$ are plotted against $t$ 
for $p=2.0$, $N=64$ and $\kappa=6$, where 
we use the block size $n=4$.
The lines are drawn to guide the eye.
\label{fig:teigen_p20}}
\end{figure}

Another interesting observation here 
concerns the dimensionality of the space.
In Fig.~\ref{fig:teigen_p20} we plot 
the expectation values of 
the eigenvalues $\lambda_{i}\left(t\right)$
of $T_{ij}\left(t\right)$
obtained for $p=2.0$, $N=64$ and $\kappa=6$.
We observe five large values near $t=0$, which indicates
the emergence of a 5d structure.
However, we emphasize that this by no means implies that 
(5+1)d space-time can also appear from the model
since one cannot take a sensible large-$N$ limit for $p=2.0$.

\section{Summary}
\label{sec:summary}


In this paper, we have addressed an important issue 
in the Lorentzian type IIB matrix model
concerning 
the IR cutoffs, which are inevitably introduced
to make the model well-defined.
In particular,
we have generalized the form of the IR cutoffs
as (\ref{eq:t_cutoff}) and (\ref{eq:s_cutoff}) with a parameter $p$,
and performed Monte Carlo simulation of the simplified model for 
various values of $p$.
From the results obtained for $p\le 1.5$,
we observe a universal behavior
for $p=1.3$, 1.4, 1.5 except near the ``boundary''.
This suggests the possibility that 
the effects of the IR cutoffs vanish
in the 
infinite-volume limit
for sufficiently large $p$.
In order to clarify this possibility, we have investigated 
the effects of the IR cutoffs directly by the SDE. 
The results show clear tendency that the IR cutoff effects 
decrease as we take the infinite-volume limit for sufficiently large $p$.

On the other hand, for $p$ as large as $2.0$, we 
observe that the number of data points in the region with
the spontaneous breaking of SO(9) symmetry does not increase
with $N$, and that large-$N$ scaling behaviors do not show up.
Combining this with the results obtained for $p\le 1.5$,
we conclude that there exists a finite range of $p$, in which
a sensible large-$N$ limit can be taken
and the results become independent of $p$.
While this range of $p$ does not 
include the value $p=1.0$ used in the previous work,
the qualitative properties of the model such as the dimensionality
of the emergent space-time and the exponential expansion
remain the same.
It is also interesting that a (5+1)d structure
is observed for $p=2.0$, where a sensible large-$N$ limit cannot be taken.

We consider that a similar conclusion holds also in 
the original Lorentzian type IIB matrix model
since the simplified model captures
the early time behaviors qualitatively.
It is therefore
important to study the original model
with various $p$
and to identify the region of $p$, in which
a universal large-$N$ limit can be taken.
We hope to address this issue in future publications.


\section*{Acknowledgements}

We
thank T.~Azuma and S.-W.~Kim for their participation
at the early stage of this work.
We are also grateful to H.~Kawai for valuable comments and 
discussions.
This research was supported by MEXT as
``Priority Issue on Post-K computer'' 
(Elucidation of the Fundamental Laws and Evolution of the Universe) 
and 
Joint Institute for Computational Fundamental Science (JICFuS).
Computations were carried out
using computational resources of the K computer 
provided by the RIKEN Advanced Institute for Computational Science 
through the HPCI System Research project (Project ID:hp150082).
The supercomputer FX10 at University of Tokyo has been used
in developing our code for parallel computing.
J.~N.\ and A.~T.\ were supported in part by Grant-in-Aid 
for Scientific Research (No.\ 23244057 and 15K05046, respectively)
from Japan Society for the Promotion of Science.

\appendix

\section{Results for the spatial SDE}
\label{sec:spatialSDE}

\begin{figure}[p]
\begin{centering}
\includegraphics[width=7cm]{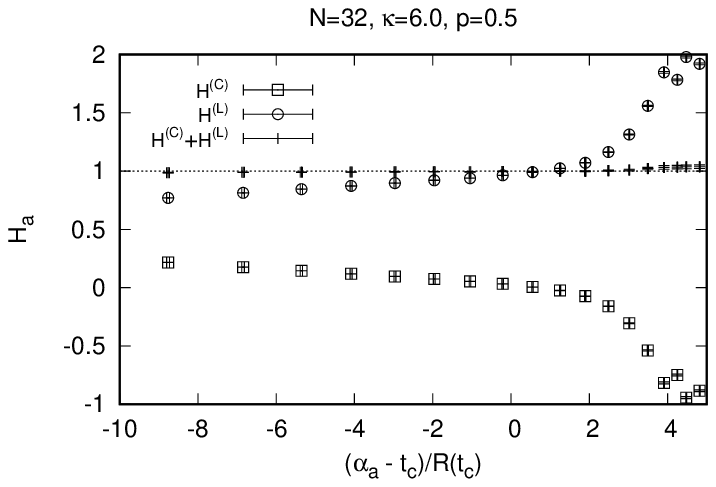}
\includegraphics[width=7cm]{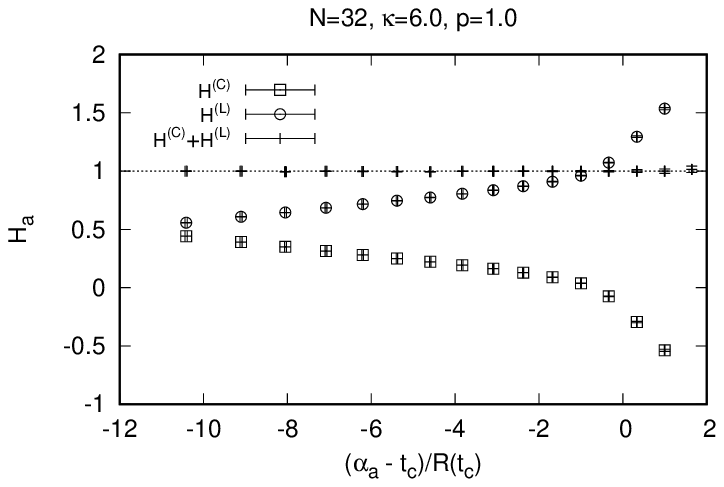}
\par\end{centering}

\centering{}
\includegraphics{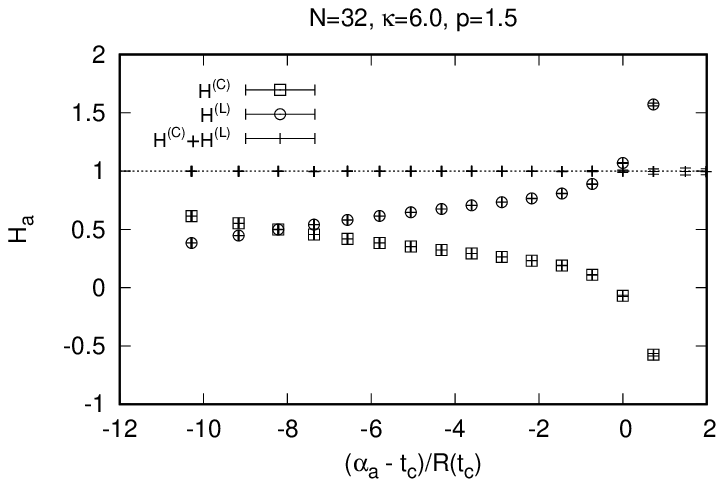}
\caption{The terms $H^{(C)}_a$ and $H^{(L)}_a$ 
in the spatial SDE (\ref{eq:sdeq_spc-3}) are plotted against 
$x=\left(\alpha_{a}-t_{\mathrm{c}}\right)/R\left(t_{\mathrm{c}}\right)$
for $p=0.5$ (Top-Left), $p=1.0$ (Top-Right) and $p=1.5$ (Bottom)
with $N=32$ and $\kappa=6$.
We also plot the sum of the two terms,
which agrees well with $1-1/N^{2}$ represented by the dotted lines.
\label{fig:sdeq_check_spc}}
\end{figure}

In this section, we present our
results for the spatial SDE \eqref{eq:sdeq_spc-3}.
In Fig.~\ref{fig:sdeq_check_spc}, 
we plot $H^{(C)}_a$, $H^{(L)}_a$ and their sum
against 
$x=\left(\alpha_{a}-t_{\mathrm{c}}\right)/R\left(t_{\mathrm{c}}\right)$
obtained for $N=32$ and $\kappa=6$ with $p=0.5,\,1.0$ and $1.5$. 
We find that 
the spatial SDE \eqref{eq:sdeq_spc-3}
is actually satisfied
at every $\alpha_{a}$. 
The effects of the IR cutoffs in the spatial direction
represented by $H^{(L)}_a$ become large towards the 
``boundary'' represented by the right-most point in these plots,
where the extent of space $R\left(t\right)$ becomes maximum.


\begin{figure}[p]
\begin{centering}
\includegraphics[width=7cm]{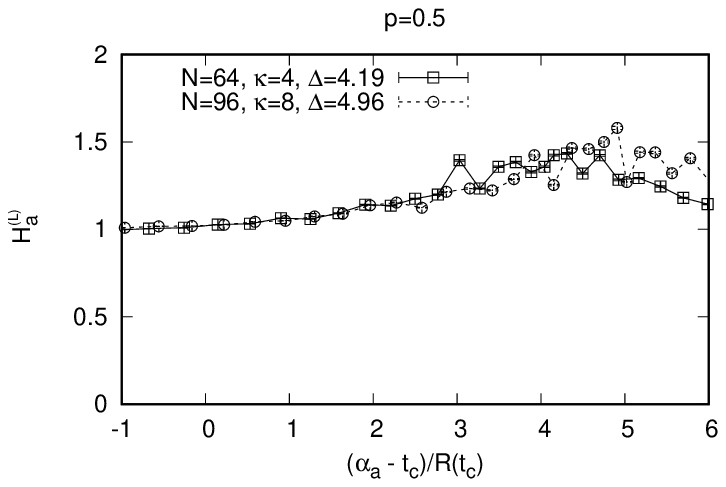}
\includegraphics[width=7cm]{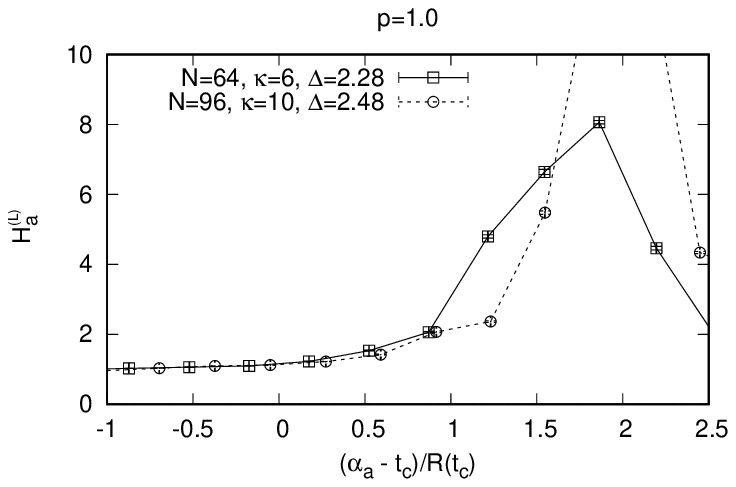}
\par\end{centering}

\centering{}
\includegraphics{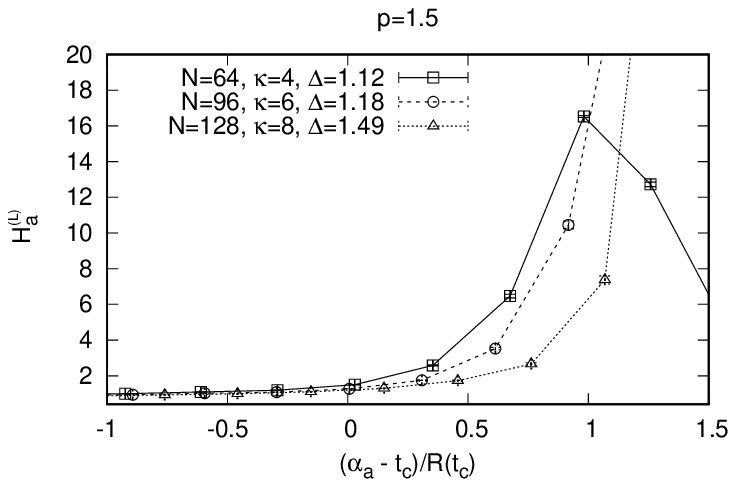}
\caption{
The IR cutoff term $H^{(L)}_a$
in the spatial SDE (\ref{eq:sdeq_spc-3}) is plotted against 
$x=\left(\alpha_{a}-t_{\mathrm{c}}\right)/R\left(t_{\mathrm{c}}\right)$
for $p=0.5$ (Top-Left), $p=1.0$ (Top-Right) and $p=1.5$ (Bottom)
with $\kappa$ and $N$
given in table \ref{tab:infinite_volume}.
The lines are drawn to guide the eye.
\label{fig:infinite_volume_spc}}
\end{figure}

In Fig.~\ref{fig:infinite_volume_spc}, we plot
$H^{(L)}_a$
against
$x=\left(\alpha_{a}-t_{\mathrm{c}}\right)/R\left(t_{\mathrm{c}}\right)$
for $p=0.5$, $p=1.0$ and $p=1.5$
with the parameters $N$ and $\kappa$
chosen as in table \ref{tab:infinite_volume}.
Let us
focus on the region in which we observe scaling behaviors
in Fig.~\ref{fig:Rt_p05}; namely $x \lesssim 4$ for $p=0.5$,
$x \lesssim 1$ for $p=1.0$ and $x \lesssim 0.6$ for $p=1.5$.
In these scaling regions,
we find for $p=0.5$ and $p=1.0$ that $H^{(L)}_a$
is more or less independent of the volume $\Delta$,
whereas for $p=1.5$, we see a clear trend showing that
it decreases with the volume $\Delta$.

\end{document}